\documentclass{amsart}[11pt]
\usepackage[margin=4cm]{geometry}
\usepackage{amsmath, amsfonts, amsthm, amssymb, mathtools}
\usepackage{mathrsfs}
\usepackage{subfig}
\usepackage{stmaryrd}
\usepackage{verbatim}
\usepackage{hyperref}
\usepackage{xcolor}
\usepackage{graphicx}
\usepackage{algorithmic, algorithm}
\usepackage{enumerate}
\usepackage{listings}
\numberwithin{equation}{section}

\usepackage[textsize=tiny]{todonotes}

\makeatletter
\def\@tocline#1#2#3#4#5#6#7{\relax
  \ifnum #1>\c@tocdepth 
  \else
    \par \addpenalty\@secpenalty\addvspace{#2}%
    \begingroup \hyphenpenalty\@M
    \@ifempty{#4}{%
      \@tempdima\csname r@tocindent\number#1\endcsname\relax
    }{%
      \@tempdima#4\relax
    }%
    \parindent\z@ \leftskip#3\relax \advance\leftskip\@tempdima\relax
    \rightskip\@pnumwidth plus4em \parfillskip-\@pnumwidth
    #5\leavevmode\hskip-\@tempdima
      \ifcase #1
       \or\or \hskip 1em \or \hskip 2em \else \hskip 3em \fi%
      #6\nobreak\relax
    \dotfill\hbox to\@pnumwidth{\@tocpagenum{#7}}\par
    \nobreak
    \endgroup
  \fi}
\makeatother

\hypersetup{colorlinks,linkcolor=blue!55!black,citecolor=blue!55!black,urlcolor=blue!55!black}

\mathtoolsset{showonlyrefs=true}
\newtheorem{theorem}{Theorem}[section]
\newtheorem{corollary}[theorem]{Corollary}
\newtheorem{lemma}[theorem]{Lemma}
\newtheorem{proposition}[theorem]{Proposition}

\theoremstyle{definition}

\newtheorem{remark}[theorem]{Remark}
\newtheorem{example}[theorem]{Example}
\newtheorem{assumption}[theorem]{Assumption}

\newcommand{\ind}{1\hspace{-2.1mm}{1}} 
\newcommand{\half}{\frac{1}{2}}

\newcommand{\CC}{\mathbb{C}}
\newcommand{\RR}{\mathbb{R}}
\newcommand{\NN}{\mathbb{N}}
\newcommand{\PP}{\mathbb{P}}

\newcommand{\EE}{\mathbb{E}}

\newcommand{\VV}{\mathbb{V}}

\newcommand{\Cc}{\mathcal{C}}

\newcommand{\Ff}{\mathcal{F}}
\newcommand{\Nn}{\mathcal{N}}
\newcommand{\Oo}{\mathcal{O}}

\newcommand{\Ss}{\mathcal{S}}
\newcommand{\Ww}{\mathcal{W}}

\newcommand{\D}{\mathrm{d}}
\newcommand{\E}{\mathrm{e}}

\newcommand{\eps}{\varepsilon}

\newcommand{\Cov}{\CC\mathrm{ov}}

\newcommand{\Af}{\mathfrak{A}}
\newcommand{\Hf}{\mathfrak{H}}
\newcommand{\af}{\mathfrak{a}}
\newcommand{\cf}{\mathfrak{c}}

\newcommand{\df}{\mathfrak{d}}

\newcommand{\rhobar}{\overline{\rho}}

\newcommand{\clock}{\tau^{R}}
\newcommand{\clockhat}{\tau^{R}}

\title{Feeling the generalised rough SABR boundary}
\author{Masaaki Fukasawa}
\address{Graduate School of Engineering Science, University of Osaka}
\email{fukasawa@sigmath.es.osaka-u.ac.jp}
\author{Antoine Jacquier}
\address{Department of Mathematics, Imperial College London}
\email{a.jacquier@imperial.ac.uk}
\author{Mugad Oumgari}
\address{Department of Mathematics, University College London}
\email{mugad.attahallah.20@ucl.ac.uk}

\date{\today}

\begin{document}

\begin{abstract}
We provide an in-depth analysis of a generalised rough version of the SABR volatility model, widely used in the financial industry.
We specifically focus on its behaviour at the origin, whether it can reach it and how fast it may do so.
\end{abstract}
\maketitle

\setcounter{tocdepth}{3}
\tableofcontents

\section{Introduction}
The stochastic alpha-beta-rho (SABR) model introduced in~\cite{hagan2002managing} has become a standard in volatility modelling. 
Its success is unusual in combining a parsimonious stochastic volatility specification with an explicit and remarkably effective approximation of implied volatility.
The model is widely used, particularly in interest-rate and foreign-exchange markets and has generated a substantial mathematical literature on implied  and stochastic volatility, and on asymptotic analysis. 
In its classical form, the forward price $S$ satisfies
\begin{equation}\label{eq:sabr_intro}
\frac{\D S_t}{S_t^\beta}
=\sigma_t \D W_t,
\qquad
\frac{\D \sigma_t}{\sigma_t}
=\nu \D B_t,
\qquad
\D \langle W,B\rangle_t=\rho \D t,
\end{equation}
where $\beta\in[0,1]$ and $\rho \in [-1,1]$. 
Thus the volatility has a dependence on the price level in the constant-elasticity-of-variance (CEV) form, with a stochastic volatility coefficient.

For \(\beta<1\), the CEV structure raises an important issue absent from the lognormal case \(\beta=1\): the process may reach zero in finite time,
and its behaviour there must be specified as part of the model. 
This is classical for the CEV model and has been studied extensively for SABR;
Chen and Yang~\cite{chen2019principle} investigated this absorption at zero and established a ``principle of not feeling the boundary'' for SABR:
although zero is attainable, 
its effect is exponentially small over short time,
which explains why short-maturity approximations derived without explicitly taking account of the boundary nevertheless remain accurate away from zero. 
It also illustrates an important distinction between whether the boundary is attainable and how likely it is to be reached over short time.

More recently, rough volatility models have emerged as an extension of classical stochastic volatility models. 
Motivated in part by the empirical observation that volatility exhibits substantially lower regularity than Brownian semimartingales~\cite{GJR}, these models replace the Brownian volatility factor by a Volterra process with local regularity described by a Hurst parameter $H<\half$. 
Besides providing parsimonious descriptions of volatility dynamics, rough volatility models naturally reproduce the pronounced maturity dependence of short-dated implied volatility skews.
The combination of rough volatility with the SABR structure has been studied from several complementary viewpoints. Rough extensions of SABR and their short-maturity asymptotics were developed in \cite{FukasawaAAA, fukasawa2021rough, FHT, FT}. These models retain  dependence of instantaneous volatility on the underlying while replacing the classical lognormal Brownian volatility factor by a rough Volterra factor. They thereby preserve the basic SABR mechanism with the effective approximation of implied volatility
while incorporating the short-time scaling characteristic of rough volatility.

The presence of rough volatility changes the boundary problem substantially. 
Standard diffusion arguments available for CEV and classical SABR cannot be applied directly, since the volatility is neither Markovian nor a semimartingale. At the same time, one expects the behaviour of the local volatility coefficient near zero to remain decisive for attainability of the boundary. This raises two natural questions: how the classical boundary classification is modified in rough volatility, and, when zero is attainable, on what scale the probability of reaching it vanishes as maturity tends to zero.

The first purpose of this paper is to answer the former question for a generalised class of rough SABR models, where the forward price process satisfies
$$
  \D S_{t}=\beta(S_{t})\sigma_{t} \D W_{t},\quad S_0>0,
$$
with~$(\sigma_{t})_{t\geq 0}$ as in~\eqref{eq:sabr_intro},
but with minimal assumptions on the local volatility function~$\beta(\cdot)$.
For this, we provide a boundary classification in terms of~$\beta(\cdot)$ only. 
This recovers the familiar CEV classification as a special case and clarifies which features of the classical boundary behaviour survive the introduction of rough volatility.
For
$\tau :=\inf\{t>0:S_t=0\}$,
we further establish a large-deviations-type asymptotic of the form
$$
\frac{T^{2H}}{\log(T)^2}
\log\PP(\tau \leq T)
\longrightarrow -I,
\qquad\text{as } T\downarrow0,
$$
with an explicitly characterised rate \(I>0\). 
This may be viewed as a rough-volatility counterpart of the no-feeling-boundary principle from~\cite{chen2019principle}. 
It not only determines whether zero is attainable but quantifies how rapidly the attainable boundary becomes invisible at short maturities. Together with the boundary classification, it provides a systematic description of the zero boundary for generalised rough SABR models.

\section{Rough SABR}\label{sec:model}
Let $(\Omega, \mathcal{F}, (\mathcal{F}_{t})_{t \geq 0}, \PP)$ be a
filtered probability space carrying a two-dimensional 
Brownian motion $(B, B^{\perp})$ and let $W := \rho B + \rhobar B^{\perp}$, $\rho \in [-1,1]$ and
$\rhobar:=\sqrt{1-\rho^2}$.
\subsection{The model}\label{sec:model}
Following~\cite{fukasawa2021rough}, the
rough SABR model with  $\beta \in (0,1)$ is defined as
\begin{equation}\label{eq:rSABR}
\frac{\D S_{t}}{S_{t}^{\beta}} = \sigma_{t} \D W_{t}, \qquad
\frac{\D\xi_{t}(s)}{\xi_{t}(s)} = \eta\sqrt{2H} (s-t)^{H - \half} \D B_{t}, \quad t \in (0,s),
\end{equation}
where $\sigma_{t} := \sqrt{\xi_{t}(t)}$,
for $\eta, S_0, \sigma_0 > 0$ and $H \in (0, \half]$.
The unique strong solution of the forward variance equation in~\eqref{eq:rSABR} reads
\begin{equation}\label{eq:xi_explicit}
  \xi_{t}(s) = \xi_0(s)\exp\left\{
\eta\sqrt{2H}\int_0^t (s-u)^{H-\half} \D B_u
- \frac{\eta^2}{2}\Big(s^{2H} - (s-t)^{2H}\Big)
  \right\}, \quad 0 \leq t \leq s.
\end{equation}
When $H = \half$,
then
$\sigma_{t}
= \xi_{t}(t)^{\half}
 = \xi_0(t)^{\half}\E^{
\frac{\eta}{2} B_{t}
- \frac{\eta^2 t}{4}}$,
so that, assuming $\log\xi_{0}$ differentiable,
$$
\frac{\D\sigma_{t}}{\sigma_{t}}
= \frac{\eta}{2}\D B_{t}
+ \half\left(\partial_{t}\log(\xi_{0}(t)) - \frac{\eta^2}{4}\right)\D t.
$$
Taking $\xi_0(t) = \sigma_{0}^2\exp\{\frac{\eta^2 t}{4}\}$ reduces to  
classical SABR~\cite{hagan2015probability,hagan2002managing} 
$\D\sigma_{t} = \frac{\eta}{2} \sigma_{t} \D B_{t}$.
\begin{remark}
In the $H=\half$ case, 
the process~$S$ is a true martingale, as proved in~\cite{lions2007correlations} for the $\beta=1$ and $\rho\leq 0$ case and in~\cite{lewis2009option} and~\cite[Proposition~2.3]{pirjol2025vix} for $\beta<1$, essentially following~\cite{sin1998complications}, when assuming absorbing boundary at the origin.
In the rough Lognormal case with $\beta=1$ and $H \in (0,\half)$, Gassiat proved the true martingale property (if and only if  $\rho\leq 0$) in~\cite{gassiat2019martingale}.
Fukasawa, Jacquier and Pannier recently~\cite{fukasawa2026weak} extended this to a more general rough local-stochastic volatility model 
(including power-law behaviour $\beta\in (0,1)$).
\end{remark}
Without assuming flat initial forward variance curve, we have
\begin{equation}\label{eq:alpha_explicit}
\sigma_{t}^2 = \xi_0(t) \exp\left\{
\eta\sqrt{2H}\int_0^t (t-u)^{H-\half} \D B_u
- \frac{\eta^2}{2} t^{2H}
\right\} =:
\xi_0(t) \exp\left\{M_{t} - \frac{\eta^2}{2} t^{2H}\right\},
\end{equation}
where 
$M_{t} := \eta\sqrt{2H}\int_0^t (t-u)^{H-\half} \D B_u$ is centred Gaussian  with
$\VV[M_{t}] = \eta^2 t^{2H}$,
\begin{align}\label{eq:c_def}
\CC(s,t) := \Cov(M_s, M_{t})
 & = \EE\left[M_{t} M_s\right] - \EE\left[M_{t}\right]\EE\left[M_s\right]\nonumber\\
 & = 2H \eta^2\EE\left[\left(\int_0^t (t-u)^{H-\half} \D B_u\right)\left(\int_0^s (s-u)^{H-\half} \D B_u\right)\right]\nonumber\\
 & = 2H \eta^2\int_0^{t\wedge s} (t-u)^{H-\half}(s-u)^{H-\half} \D u\nonumber\\
& = 2H \eta^2 \int_0^s (s-u)^{H-\half}(t-u)^{H-\half} \D u,\nonumber\\
 & = \frac{2H\eta^2}{H + \half} s^{H+\half}t^{H-\half} _{2}\!\mathrm{F}_{1} \left(\half-H,1;H+\frac{3}{2};\frac{s}{t}\right),
 \quad\text{for }s \leq t,
\end{align}
extended symmetrically by $\CC(s,t) = \CC(t,s)$
and~$_{2}\!\mathrm{F}_{1}$ the Hypergeometric function~\cite[\S15]{abramowitz1966handbook}.

\subsection{Total variance and the flat initial forward variance curve assumption}
We slightly digress from the main objective to discuss
assumptions on the initial forward variance curve~$\xi_{0}(\cdot)$.
It is often assumed flat in the literature,
though this is rarely argued or analysed.
Introduce the total variance
\begin{equation}\label{eq:TotalVar}
\Af_{T} := \int_{0}^{T}\sigma_{s}^2 \D s,
\end{equation}
and the $[0,\infty]$-valued quantities
\begin{align*}
\mathcal{V}_{T} & := \int_{0}^{T}\int_{0}^{T}\xi_0(t)\xi_0(s)\left(\E^{\CC(s,t)}-1\right) \D s \D t,\\
\mathcal{J}_{T} & := \int_{0}^{T} t^{H-\half}\xi_0(t)\left(\int_0^t s^{H+\half}\xi_0(s) \D s\right)\D t.
\end{align*}
While one might expect $\VV[\Af_{T}]=\mathcal{V}_{T}$, 
this does not always hold
as the following shows.
 
\begin{proposition}\label{prop:AT}
Let $H\in(0,\half]$ and  $\xi_0:[0,\infty)\to[0,\infty)$ be measurable.
\begin{enumerate}[(i)]
\item The following double inequality always holds:
$$
\frac{4H\eta^2}{H+\half} \mathcal{J}_{T} \le \mathcal{V}_{T} \le 2\eta^2\E^{\eta^2T^{2H}} \mathcal{J}_{T}.
$$
so that
$\mathcal{V}_{T}<\infty$ if and only if $\mathcal{J}_{T}<\infty$, and both hold whenever
$\int_{0}^{T} t^H\xi_0(t) \D t<\infty$.
\item The statements
$\xi_0\in L^1([0,T])$, $\Af_{T}\in L^1$ and $\Af_{T}\in L^2$
are equivalent.
In this case $\VV[\Af_{T}]=\mathcal{V}_{T}$. 
Otherwise $\Af_{T}=+\infty$ almost surely.
\item Assume there exist $\alpha\in\RR$,
$\delta\in(0,T]$ and $0<c\le c'$ such that $c t^{-\alpha}\le\xi_0(t)\le c' t^{-\alpha}$ for all
$t\in(0,\delta]$, and $\xi_0\in L^1([\delta,T])$. Then
$$
\Af_{T}\in L^2\iff\alpha<1,
\qquad\text{and}\qquad
\mathcal{V}_{T}<\infty\iff \mathcal{J}_{T}<\infty\iff\alpha<1+H.
$$
\end{enumerate}
\end{proposition}
\begin{remark}
When $\alpha \in [1,1+H)$, $\mathcal{V}_{T}$ is finite 
but not~$\Af_{T}$,
so $\mathcal{V}_{T} \ne \VV[\Af_{T}]$ in general.
\end{remark}
\begin{proof}
Set $\kappa:=\frac{2H}{H+\half}$.
For $0<s\le t$, the substitution $u=sv$ gives
$$
\CC(s,t) = 2H\eta^2 s^{H+\half}t^{H-\half}
\int_0^1(1-v)^{H-\half}\left(1-\frac{s}{t}v\right)^{H-\half}\D v.
$$
Since $1-v\le 1-\frac{s}{t}v\le 1$ on $[0,1]$ and $H-\half\le 0$, the integrand lies between
$(1-v)^{H-\half}$ and $(1-v)^{2H-1}$, so the integral lies in
$[\frac{1}{H+\half},\frac{1}{2H}]$. Hence
\begin{equation}\label{eq:Cbounds}
\kappa\eta^2 s^{H+\half}t^{H-\half}\le \CC(s,t)\le\eta^2 s^{H+\half}t^{H-\half},
\qquad 0<s\le t.
\end{equation}
Since
$s^{H+\half}t^{H-\half} = s^Ht^H (\frac{s}{t})^{\half} \le s^Ht^H$ for $s\le t$, symmetry of $\CC$ yields
$0\le \CC(s,t)\le\eta^2s^Ht^H\le \eta^2 T^{2H}$ for all $s,t\in[0,T]$.
Now, for $x\in[0,\eta^2 T^{2H}]$, $\E^x-1=\int_0^x\E^y \D y\in[x, x\E^{\eta^2 T^{2H}}]$. 
By symmetry
of the integrand and Tonelli's theorem,
$$
\mathcal{V}_{T}=2\int_{0}^{T}\xi_0(t)\int_0^t\xi_0(s)\left(\E^{\CC(s,t)}-1\right) \D s \D t.
$$
For $0<s\le t\le T$, \eqref{eq:Cbounds} and the previous bounds give
$\kappa\eta^2s^{H+\half}t^{H-\half}\le \E^{\CC(s,t)}-1\le \E^{\eta^2 T^{2H}}\eta^2s^{H+\half}t^{H-\half}$, and
integrating yields $2\kappa\eta^2\mathcal{J}_{T}\le\mathcal{V}_{T}\le2\eta^2\E^{\eta^2 T^{2H}}\mathcal{J}_{T}$. For the last claim, let
$f(t):=t^H\xi_0(t)$. 
Since $s^{H+\half}t^{H-\half}\le s^Ht^H$ for $s\le t$, the same symmetry
argument gives $\mathcal{J}_{T}\le\int_{0}^{T}f(t)\int_0^tf(s) \D s \D t=\half(\int_{0}^{T} f(t)\D t)^2$.
 
Regarding~(ii), 
applying Tonelli's theorem and the previous bound,
$\EE[\Af_{T}]=\int_{0}^{T}\xi_0(t)\D t$ and
$\EE[\Af_{T}^2]=\int_{0}^{T}\int_{0}^{T}\xi_0(s)\xi_0(t)\E^{\CC(s,t)} \D s \D t$, so that, since
$0\le \CC(\cdot,\cdot)\le \eta^2 T^{2H}$,
$$
\left(\int_{0}^{T}\xi_0(t)\D t\right)^2\le\EE\left[\Af_{T}^2\right]\le \E^{\eta^2 T^{2H}}\left(\int_{0}^{T}\xi_0(t)\D t\right)^2.
$$
Thus $\xi_0\in L^1([0,T])$ implies $\Af_{T}\in L^2$, which implies $\Af_{T}\in L^1$, which in turn implies
$\int_{0}^{T}\xi_0=\EE[\Af_{T}]<\infty$. In this case all integrals above are finite, and
$\VV[\Af_{T}]=\EE[\Af_{T}^2]-\EE[\Af_{T}]^2=\mathcal{V}_{T}$. 
Otherwise,
since~$M$ has continuous paths, then
$\Af_{T}\geq \exp\{\inf_{t\in[0,T]}M_{t}-\frac{\eta^2}{2}T^{2H}\}\int_{0}^{T}\xi_0(t)\D t = +\infty$ almost surely.
 
We now prove~(iii).
Under the assumptions, $\int_{0}^{T}\xi_0(t)\D t$ is finite if and only if
$\int_0^\delta t^{-\alpha}\D t$ is, namely $\alpha<1$; 
the first equivalence follows from (ii).
If $\alpha < 1+H$, then
$\int_{0}^{T}t^H\xi_0(t) \D t\le c'\int_0^\delta t^{H-\alpha}\D t+T^H\int_\delta^T\xi_0(t)\D t<\infty$, so
$\mathcal{J}_{T}<\infty$ and $\mathcal{V}_{T}<\infty$ by (i). If $\alpha\ge1+H$, restricting both integrals in
$\mathcal{J}_{T}$ to $(0,\delta]$ gives
$$
\mathcal{J}_{T}\ge c^2\int_0^\delta t^{H-\half-\alpha}\left(\int_0^ts^{H+\half-\alpha} \D s\right)\D t.
$$
If $\alpha\ge H+\frac{3}{2}$, the inner integral is infinite for every $t\in(0,\delta]$. If
$H+1\le\alpha<H+\frac{3}{2}$, it equals $t^{H+\frac{3}{2}-\alpha}/(H+\frac{3}{2}-\alpha)$, and the right-hand side
is a positive multiple of $\int_0^\delta t^{2H+1-2\alpha}\D t=+\infty$, since $2H+1-2\alpha\le-1$.
In both cases $\mathcal{J}_{T}=\infty$, hence $\mathcal{V}_{T}=\infty$ by (i).
\end{proof}

We are now interested in~$\Af_{T}$ as $T\uparrow\infty$. 
Clearly $T\mapsto\Af_{T}$ is non-decreasing almost surely and the limit $\Af_{\infty} := \lim_{T\uparrow\infty}\Af_{T}$ always exists in $[0,\infty]$.
From~\eqref{eq:alpha_explicit}, define
$\psi(t) := \log(\xi_{0}(t)) - \frac{\eta^2}{2}t^{2H}$,
so that $\sigma_{t}^2 = \exp\{\psi(t) + M_{t}\}$.
\begin{proposition}\label{prop:Af_infty}
If $\xi_{0} \notin L^{1}([0,T])$ for some $T>0$, then $\Af_{t} = \infty$ almost surely for all $t\geq T$.
Assume now $\xi_{0} \in L^{1}([0,T])$ for all $T>0$ and let $\ell(t) := t^{H}\sqrt{\log(\log(t))}$ for $t>\E$.
\begin{enumerate}[(i)]
\item If there exist $a\in\RR$, $c>0$ and $t^{*}>0$ such that, for all $t\geq t^{*}$,
\begin{equation}\label{eq:Cond_xi0}
\xi_{0}(t) \geq c\exp\left\{\frac{\eta^2}{2}t^{2H} - at^{H}\right\},
\end{equation}
then $\Af_{\infty} = \infty$ almost surely;
\item If $\int^{\infty}\exp\{\psi(t) - c\ell(t)\}\D t = \infty$ for some $c>\eta\sqrt{2}$ then $\Af_{\infty}=\infty$ almost surely;
\item If $\int^{\infty}\exp\{\psi(t) + c\ell(t)\}\D t < \infty$ for some $c>\eta\sqrt{2}$ then $\Af_{\infty}<\infty$ almost surely;
\end{enumerate}
\end{proposition}

The following corollary provides conditions easier to check:
\begin{corollary}\label{cor:Af_infty}
Assume $\xi_{0} \in L^{1}([0,T])$ for all $T>0$.
\begin{enumerate}[(i)]
\item If $\liminf_{t\uparrow\infty}t^{-2H}\log(\xi_{0}(t)) > \half\eta^2$, then $\Af_{\infty} = \infty$ almost surely;
\item If $\limsup_{t\uparrow\infty}t^{-2H}\log(\xi_{0}(t)) < \half\eta^2$, then $\Af_{\infty} <\infty$ almost surely.
\end{enumerate}
\end{corollary}
\begin{remark}
In classical $H=\half$ SABR,
$\xi_{0}(t) = \sigma_{0}^{2}\exp\{\frac{\eta^2 t}{4}\}$, 
so $\limsup_{t\uparrow\infty}
t^{-1}\log(\xi_{0}(t)) = \frac{\eta^2}{4}<\half\eta^2$
and $\Af_{\infty}$ is finite almost surely.
\end{remark}

\begin{proof}[Proof of Proposition~\ref{prop:Af_infty}]
The $\xi_{0} \notin L^{1}([0,T])$ statement is obvious.
Assume now that $\xi_{0} \in L^{1}([0,T])$ for all $T>0$.
Regarding~(i), using the fact that the transformed process $\widetilde{M}_{t}:=\E^{-tH}M_{\E^{t}}$ is a continuous, stationary centered Gaussian process
with $\VV[\widetilde{M}_{0}] = \eta^2$
and decaying covariance from~\eqref{eq:Cbounds}, and hence ergodic, then, assuming~\eqref{eq:Cond_xi0}, for $t\geq t^{*}$,
$\psi(t) = \log(\xi_{0}(t))  -\frac{\eta^2}{2}t^{2H}
\geq \log(c) - at^{H}$
and
$$
\sigma_{t}^{2}
= \E^{\psi(t) + M_{t}}
\geq c\exp\left\{- at^{H} + t^{H}\widetilde{M}_{\log(t)}\right\}
= c\exp\left\{t^{H}\left(\widetilde{M}_{\log(t)}-a\right)\right\}
\geq c\ind_{\left\{\widetilde{M}_{\log(t)}\geq a\right\}},
$$
and therefore
\begin{align*}
\Af_{\infty}
= \Af_{t^{*}} + \int_{t^{*}}^{\infty}\sigma_{t}^{2}\D t  
\geq c\int_{t^{*}}^{\infty}\ind_{\left\{\widetilde{M}_{\log(t)}\geq a\right\}}\D t 
 & = c\int_{\log(t^{*})}^{\infty}\E^{u}\ind_{\left\{\widetilde{M}_{u}\geq a\right\}}\D u\\
 & \geq c\int_{\log(t^{*})\vee 0}^{\infty}\ind_{\left\{\widetilde{M}_{u}\geq a\right\}}\D u.
\end{align*}
Since $\widetilde{M}$ is ergodic, Birkhoff's ergodic theorem~\cite[Theorem~25.6]{kallenberg1997foundations} implies that
$$
\lim_{t\uparrow\infty}\frac{1}{t}\int_{0}^{t}\ind_{\left\{\widetilde{M}_{u}\geq a\right\}}\D u
 = \PP\left(\widetilde{M}_{0} \geq a\right)>0,
$$
and therefore $\Af_{\infty} = \infty$ almost surely, proving~(i).
From~\cite[Theorem~1.1]{orey1972growth}
(with $\beta_1 = \beta_2 = \beta_3 = 2H$ and $Q(t) = v(t) = \half \eta^2 t^{2H}$), 
the law of the iterated logarithm reads
$\limsup_{t\uparrow\infty}\frac{M_{t}}{\ell(t)} = \eta\sqrt{2}$.
Note that Orey's argument is stated for continuous Gaussian processes with stationary increments, but the proof does not actually requires this 
(and it in fact does not hold here) and we refer the reader to~\cite{arcones1995law,oodaira1973law} where this assumption is relaxed.
Therefore there exists an almost surely finite random time~$\widetilde{\tau}$ such that
$M_{t} \leq c\ell(t)$ for any $c> \eta\sqrt{2}$
for all $t\geq \widetilde{\tau}$.
Thus, for $t\geq \widetilde{\tau}$,
$$
\E^{\psi(t) - \eta\sqrt{2}\ell(t)}
\leq \sigma_{t}^{2}
= \E^{\psi(t) + M_{t}}
\leq \E^{\psi(t) + \eta\sqrt{2}\ell(t)},
$$
using a symmetric law of the iterated logarithm argument for~$-M$ to obtain a $\lim\inf$ statement.
In the setting of~(ii), 
$
\Af_{\infty} = \int_{0}^{\infty}\sigma_{t}^{2}\D t
\geq \int_{\widetilde{\tau}}^{\infty}\E^{\psi(t) - \eta\sqrt{2}\ell(t)}\D t
\geq \int_{\widetilde{\tau}}^{\infty}\E^{\psi(t) - c\ell(t)}\D t = \infty
$
almost surely for any $c>\eta\sqrt{2}$.
In the setting of~(iii),
for any $c>\eta\sqrt{2}$, almost surely,
$$
\Af_{\infty} = \int_{0}^{\infty}\sigma_{t}^{2}\D t
= \Af_{\widetilde{\tau}} + \int_{\widetilde{\tau}}^{\infty}\sigma_{t}^{2}\D t
\leq \Af_{\widetilde{\tau}}
+ \int_{\widetilde{\tau}}^{\infty}\E^{\psi(t) + \eta\sqrt{2}\ell(t)}\D t
\leq \Af_{\widetilde{\tau}}
+ \int_{\widetilde{\tau}}^{\infty}\E^{\psi(t) + c\ell(t)}\D t
< \infty.
$$
\end{proof}

\subsection{Change of variables and Bessel structure}\label{sec:Bessel_Structure}

Following~\cite[Section~2]{chen2019principle}, define
\begin{equation}\label{eq:X_def}
X_{t} := \frac{S_{t}^{1-\beta}}{1-\beta},
\end{equation}
and let $\tau := \inf\{t \geq 0 : S_{t} = 0\}
= \inf\{t \geq 0 : X_{t} = 0\}$ denote the first hitting time of the origin.  
It\^{o}'s formula gives
\begin{equation}\label{eq:X_SDE}
\D X_{t}
 = \sigma_{t}\D W_{t}
  + \frac{\gamma\sigma_{t}^2}{X_{t}} \D t,
\end{equation}
for $t\leq \tau$, and we assume an absorbing boundary at the origin, where
\begin{equation}\label{eq:gamma}
\gamma
:= -
 \frac{\beta}{2(1-\beta)}<0.
 \end{equation}
By Dambis-Dubins-Schwarz~\cite[Chapter~5, Theorem~1.6]{revuz2013continuous}, there exists a standard Brownian motion~$\widetilde{W}$ (on a possibly enlarged probability space, since~$\Af_{\infty}$ may be finite)
such
that $\int_0^t \sigma_s \D W_s = \widetilde{W}_{\Af_{t}}$.
Then $R_{t} := X_{\Af^{\leftarrow}_{t}}$, for $t<\Af_{\tau}$, 
satisfies the SDE
\begin{equation}\label{eq:bessel_SDE}
  \D R_{t} = \D\widetilde{W}_{t}
   + \gamma  \frac{\D t}{R_{t}},
  \qquad \text{starting from }R_{0} = x_{0} = \frac{S_0^{1-\beta}}{1-\beta} > 0,
\end{equation}
which is a Bessel process of dimension
$\delta := \frac{1-2\beta}{1-\beta} < 1$,
hitting the origin almost surely since $\delta < 2$ for $\beta \in (0,1)$
(we refer the reader to~\cite[Chapter~6]{jeanblanc2009mathematical} for details of such computations).
More precisely, since $X_{\Af^{\leftarrow}}$ is only defined on $[0,\Af_{\infty})$, 
we define~$R$ on the whole of $[0,\infty)$ as $R:=\frac{\widetilde{S}^{1-\beta}}{1-\beta}$,
with~$\widetilde{S}$ the time-homogeneous diffusion from Lemma~\ref{lem:time_change},
so that $X_{t\wedge\tau} = R_{\Af_{t}}$ for all $t\geq 0$
and
(see also~\eqref{eq:hitting_equivalence})
\begin{equation}\label{eq:hitting_equiv}
\tau \leq T
  \text{ if and only if }
  \clock \leq \Af_{T},
\qquad\text{with }
  \clock := \inf\{t \geq 0 : R_{t} = 0\}.
\end{equation}

Recall the following~\cite[\S6.2.3]{jeanblanc2009mathematical},
with~$\Gamma(\cdot,\cdot)$  the incomplete Gamma function~\cite[\S6.5]{abramowitz1966handbook}:
\begin{lemma}\label{lem:bessel_cdf}
Let $\{R_{t}\}_{t \geq 0}$ solve~\eqref{eq:bessel_SDE} with absorbing boundary at zero. For any $m>0$,
\begin{equation}\label{eq:bessel_hit_prob}
\PP\left(\clock \leq m\right)
= \frac{\Gamma\left(1 - \frac{\delta}{2}, 
\frac{x_{0}^2}{2m}\right)}{\Gamma\left(1 -
\frac{\delta}{2}\right)}.
\end{equation}
\end{lemma}

\section{Main results}
\subsection{Reaching the origin}
We consider a generalised version of~\eqref{eq:rSABR}, namely
\begin{equation}\label{eq:S_SDE}
  \D S_{t}=\beta(S_{t})\sigma_{t} \D W_{t},\quad S_0>0,
\end{equation}
where the local volatility component~$\beta(\cdot)$ is left unspecified and so is the volatility process~$\sigma$.
We assume that
$\beta\in \Cc^\infty((0,\infty) \to (0,\infty))$ and~$\sigma$ is a continuous and strictly positive almost surely process adapted to a large filtration generating~$W$ and potentially another correlated Brownian motion~$B$ (with its filtration $(\Ff_{t}^{B})_{t\geq 0}$), and let
$\tau:=\inf\{t\ge0:S_{t}=0\}$ as before.
We are chiefly interested in the case where~$\tau$ is finite, 
for which the following result provides a clean characterisation.
We denote
$$
\Sigma := \int_{0}^{1}\frac{x}{\beta(x)^2} \D x.
$$

\begin{theorem}\label{thm:feller}
Under~\eqref{eq:S_SDE}, the following hold:
\begin{enumerate}[(i)]
\item if $\Sigma = \infty$, 
then $\PP(\tau<\infty)=0$;
\item if $\Sigma < \infty$, 
then $\tau<\infty$ almost surely conditionally on $\{\Af_{\infty}=\infty\}$
and $\PP(\tau<\infty)\geq \PP(\Af_{\infty}=\infty)$.
\end{enumerate}
Consequently, if $\PP(\tau<\infty)>0$ then~$\Sigma$ is finite, 
and the converse holds as soon as $\PP(\Af_{\infty}=\infty)>0$.
\end{theorem}

We now consider the rough SABR~\eqref{eq:rSABR}, 
where $\beta(s) = s^{\beta}$ with $\beta\in (0,1)$.
Then
$\Sigma = \frac{1}{2(1-\beta)}$ and Theorem~\ref{thm:feller}(ii) applies.
In the case of a flat forward variance curve however, 
$\Af_{\infty}$ is finite almost surely by Corollary~\ref{cor:Af_infty}(ii),
and Theorem~\ref{thm:feller} is uninformative, but the following holds:
\begin{lemma}\label{lem:rSABR_access}
In rough SABR~\eqref{eq:rSABR} with $\beta\in(0,1)$, $\rho \in (-1,1)$,
$H\in(0,\half]$, 
and~$\xi_{0}$ strictly positive with $\xi_{0}\in L^{1}([0,T])$ for all $T>0$.
Then, for all $T>0$,
with~$\Phi$ the Gaussian cdf,
$$
\PP(\tau\leq T)
\geq \EE\left[\Phi\left(-\frac{x_{0}+\rho V_{T}}{\rhobar\sqrt{\Af_{T}}}\right)\right] > 0.
$$
\end{lemma}
\begin{proof}
Fix $T>0$. 
Then $\Af_{T}$ is strictly positive and finite almost surely by Proposition~\ref{prop:AT}(ii).
Write $V_{T}:=\int_0^T\sigma_s\D B_s$, 
$V^\perp_{T}:=\int_0^T\sigma_s\D B^\perp_s$, 
and let~$X$ be as in~\eqref{eq:X_def}.
On $\{\tau>T\}$, the drift in~\eqref{eq:X_SDE} is non-positive on $[0,T]$, 
so 
$0 < X_{T} \leq x_{0} + \rho V_{T} + \rhobar V^\perp_{T}$
hence
$\{x_{0} + \rho V_{T} + \rhobar V^\perp_{T}\leq 0\}\subseteq\{\tau\leq T\}$.
Since~$\sigma$ is $\Ff^{B}$-adapted, 
then, conditionally on~$\Ff^{B}$, $V^\perp_{T}$ is centred Gaussian 
with variance~$\Af_{T}$ and the lemma follows.
\end{proof}

\subsection{Feeling the rough SABR boundary}

We now consider the rough SABR~\eqref{eq:rSABR} $\beta(s) = s^{\beta}$, with $\beta\in (0,1)$;
if the forward variance curve is such that $\Af_{\infty}=\infty$, 
then all statements in Theorem~\ref{thm:feller} apply; 
indeed, the integral in Theorem~\ref{thm:feller}(ii) thus reads
$$
\int_{0}^{1}\frac{x}{x^{2\beta}}\D x
 = \frac{1}{2(1-\beta)},
$$
which is finite.
The main result is the following theorem:
\begin{theorem}\label{thm:feeling_bound}
Assume 
$\xi_{0}(\cdot) \in \Cc([0,\infty)\to(0,\infty))$ and $\rho \in (-1,1)$.
The following behaviour holds for~\eqref{eq:rSABR} for any $H\in (0,\half)$:
\begin{equation}\label{eq:lim_theorem}
\lim_{T\downarrow 0}\frac{T^{2H}}{\log(T)^2}\log\PP(\tau \leq T)
= -\frac{(1-2H)^2}{2\eta^2}.
\end{equation}
\end{theorem}

\subsection{Extension of Theorem~\ref{thm:feeling_bound}}

The power-law form in~\eqref{eq:rSABR} is in fact not essential for Theorem~\ref{thm:feeling_bound}.
Consider~\eqref{eq:S_SDE},
with~$\sigma$ from~\eqref{eq:rSABR}-\eqref{eq:alpha_explicit}.

\begin{assumption}\label{ass:beta}
The function $\beta\in\Cc^{\infty}((0,\infty)\to(0,\infty))$ is non-decreasing and
$\varphi(s) := \int_{0}^{s}\frac{\D y}{\beta(y)}$
is finite for all $s>0$.
\end{assumption}

\begin{theorem}\label{thm:feeling_bound_general}
Under Assumption~\ref{ass:beta},
with $\rho \in (-1,1)$,
for any $H\in(0,\half)$,
$\xi_{0}\in\Cc([0,\infty)\to(0,\infty))$,
the limit~\eqref{eq:lim_theorem} still holds.
\end{theorem}

\begin{remark}\label{rem:general_beta}
\
\begin{enumerate}[(i)]
\item Assumption~\ref{ass:beta} holds for $\beta(s)=s^{\beta}$ with $\beta\in[0,1)$
for which $\varphi(s) = \frac{s^{1-\beta}}{1-\beta}$,
so Theorem~\ref{thm:feeling_bound_general}
extends Theorem~\ref{thm:feeling_bound}.
\item The limit does not depend on~$\beta(\cdot)$, $\rho$, $S_{0}$, $\xi_{0}(\cdot)$.
\item Since $\frac{1}{\beta}$ is non-increasing,
then $\varphi(x)\geq \frac{x}{\beta(x)}$ for all $x>0$, and
$$
\int_{0}^{1}\frac{x}{\beta(x)^2}\D x
\leq \int_{0}^{1}\varphi(x)\varphi'(x)\D x = \frac{\varphi(1)^2}{2}<\infty,
$$
so that the finiteness criterion from Theorem~\ref{thm:feller} is satisfied.
\item We only use monotonicity of~$\beta(\cdot)$ to prove the lower bound.
\end{enumerate}
\end{remark}

\section{Proofs}

\subsection{Proof of Theorem~\ref{thm:feller}}
Before diving into the proof of  Theorem~\ref{thm:feller}, we require some intermediate results.
We start with the time-homogeneous version of~\eqref{eq:S_SDE}:
\begin{equation}\label{eq:Stilde_hom}
\D \widetilde{S}_{t}=\beta(\widetilde{S}_{t}) \D W_{t},
\qquad \widetilde{S}_0=s_0>0,
\end{equation}
and recall the quantities (simplified from~\cite[\S15, Section~6]{karlin1981second} as~\eqref{eq:Stilde_hom} is driftless), for any $c>0$,
$s(x):=1$,
$\Ss(x):= x-c$
and $m(x):=\beta(x)^{-2}$.
With
\begin{equation}\label{eq:N_and_L}
\Sigma_{c}:=\int_0^c[\Ss(x)-\Ss(0)]m(x) \D x
 = \int_0^c\frac{x}{\beta(x)^2}\D x
\quad\text{and}\quad
\mathcal{M}_{c}:=\int_0^c m(x) \D x = \int_0^c\frac{\D x}{\beta(x)^2},
\end{equation}
Feller's classification~\cite[\S15, Section~6]{karlin1981second} reads as follows:

\begin{theorem}[Feller classification at the origin for~\eqref{eq:Stilde_hom}]
\label{thm:feller_hom}
  \begin{center}
\renewcommand{\arraystretch}{1.4}
\begin{tabular}{lllc}
\hline
Classification & $\Sigma_{c}<\infty$ & $\mathcal{M}_{c}<\infty$ &
Accessible from $(0,\infty)$\\
\hline
Regular   & Yes & Yes & Yes\\
Exit& Yes & No  & Yes\\
Entrance  & No  & Yes & No\\
Natural   & No  & No  & No\\
\hline
\end{tabular}
  \end{center}
\end{theorem}

With~$\beta(\cdot)$ continuous and strictly positive on $(0,\infty)$, 
\cite[Section~5.5.C]{karatzas1998brownian} applies:
since $\Ss(0)=-c>-\infty$ and $\Ss(\infty)=\infty$, 
then by~\cite[Proposition~5.5.22(b)]{karatzas1998brownian}, 
$\widetilde{S}$ does not explode to infinity and converges to zero as time increases to its exit time from $(0,\infty)$, 
thus equal to~$\tau^{\widetilde{S}}$,
and therefore Feller's classification yields the following:
\begin{lemma}\label{lem:dichotomy}
If $\Sigma_{c}<\infty$ then $\PP(\tau^{\widetilde{S}}<\infty)=1$;
if $\Sigma_{c}=\infty$ then $\PP(\tau^{\widetilde{S}}=\infty)=1$.
\end{lemma}

\begin{example}\label{ex:CEV}
With the CEV formulation $\beta(x)=x^\beta$ with $\beta\ge 0$, then
$\int_0^c y/\beta(y)^2 \D y=\int_0^c y^{1-2\beta} \D y
=c^{2-2\beta}/(2-2\beta)$ is finite if and only if $\beta<1$.
The origin is accessible for $\beta\in[0,1)$ and inaccessible for
$\beta\ge1$.  
When $\beta=1$ (lognormal case),
the boundary is inaccessible,
matching~\cite[Proposition~2.3]{pirjol2025vix}.
\end{example}

We now show that the criterion $\Sigma<\infty$ (Theorem~\ref{thm:feller}(ii)) for~\eqref{eq:Stilde_hom} is also the correct one for the generalised rough SABR model~\eqref{eq:S_SDE}, 
up to the behaviour of~$\Af_{\infty}$.

\begin{lemma}\label{lem:time_change}
On a possibly enlarged probability space, there exist a Brownian motion~$\widetilde{W}$ and a solution~$\widetilde{S}$ to~\eqref{eq:Stilde_hom} driven by~$\widetilde{W}$, 
starting from~$s_{0}$ and absorbed at the origin, 
such that $S_{t\wedge\tau} = \widetilde{S}_{\Af_{t}}$ for all $t\geq 0$.
\end{lemma}

\begin{proof}
Since~$\sigma$ is continuous and strictly positive, 
$t\mapsto\Af_{t}$ is continuous and strictly increasing and is the quadratic variation of the continuous local martingale $N:=\int_{0}^{\cdot}\sigma_{u}\D W_{u}$.
By the Dambis-Dubins-Schwarz theorem~\cite[\S~V, Theorem~1.6]{revuz2013continuous}
(and~\cite[\S~V, Theorem~1.7]{revuz2013continuous}
 upon enlarging the probability space, needed when~$\Af_{\infty}$ is finite with positive probability),
there exists a Brownian motion~$\widetilde{W}$, 
with respect to a filtration for which each~$\Af_{t}$ is a stopping time, 
such that $N_{t} = \widetilde{W}_{\Af_{t}}$ for all $t\geq 0$.
Since~$\beta$ is locally Lipschitz on $(0,\infty)$, 
then~\eqref{eq:Stilde_hom} driven by~$\widetilde{W}$ 
does not explode to infinity (as explained just before Lemma~\ref{lem:dichotomy})
and
admits a unique strong solution~$\widetilde{S}$ up to 
$\tau^{\widetilde{S}}:=\inf\{t\geq 0:\widetilde{S}_{t}=0\}$
and we set $\widetilde{S}_{t}:=0$ for $t\geq\tau^{\widetilde{S}}$.
The time-change formula for stochastic integrals~\cite[\S~V, Proposition~1.5]{revuz2013continuous} yields
\begin{align*}
\widetilde{S}_{\Af_{t}}
 = S_{0} + \int_{0}^{\Af_{t}}\beta(\widetilde{S}_{r})\ind_{\{r<\tau^{\widetilde{S}}\}}\D\widetilde{W}_{r}
 & = S_{0} + \int_{0}^{t}\beta(\widetilde{S}_{\Af_{u}})\ind_{\{\Af_{u}<\tau^{\widetilde{S}}\}}\D N_{u}\\
 & = S_{0} + \int_{0}^{t}\beta(\widetilde{S}_{\Af_{u}})\ind_{\{\Af_{u}<\tau^{\widetilde{S}}\}}\sigma_{u}\D W_{u},
\end{align*}
for all $t\geq 0$,
so that $\widetilde{S}_{\Af}$ solves~\eqref{eq:S_SDE} up to~$\tau^{\widetilde{S}}$, 
where it is absorbed.
Since~$\beta(\cdot)$ is locally Lipschitz, 
pathwise uniqueness holds for~\eqref{eq:S_SDE} up to~$\tau^{\widetilde{S}}$
and $S_{t\wedge\tau} = \widetilde{S}_{\Af_{t}}$ for all $t\geq 0$.
\end{proof}

\begin{lemma}\label{lem:bijection}
With~$\widetilde{S}$ and~$\tau^{\widetilde{S}}$ from Lemma~\ref{lem:time_change}, then
\begin{equation}\label{eq:hitting_equivalence}
\{\tau\leq T\} = \left\{\tau^{\widetilde{S}}\leq \Af_{T}\right\}
\quad\text{for all }T\geq 0,
\qquad\text{and}\qquad
\{\tau<\infty\} = \left\{\tau^{\widetilde{S}}< \Af_{\infty}\right\}.
\end{equation}
\end{lemma}

\begin{proof}
Let $T\geq 0$.
Since~$\widetilde{S}$ is absorbed at the origin, 
Lemma~\ref{lem:time_change} implies that 
$S_{T\wedge\tau}=0$ if and only if $\widetilde{S}_{\Af_{T}}=0$, 
that is $\tau\leq T$ if and only if $\tau^{\widetilde{S}}\leq \Af_{T}$.
Since~$\Af$ is strictly increasing to~$\Af_{\infty}$, 
then 
$\{\tau<\infty\} = \bigcup_{T\in\NN}\{\tau^{\widetilde{S}}\leq \Af_{T}\} = \{\tau^{\widetilde{S}}< \Af_{\infty}\}$.
\end{proof}

We can finally move on to the proof of Theorem~\ref{thm:feller}:

\begin{proof}[Proof of Theorem~\ref{thm:feller}]
Let~$\widetilde{S}$ and~$\tau^{\widetilde{S}}$ be as in Lemma~\ref{lem:time_change}.
Since~$\widetilde{S}$ solves~\eqref{eq:Stilde_hom}, 
Lemma~\ref{lem:dichotomy} applies to~$\tau^{\widetilde{S}}$.
If $\int_{0}^{1} \frac{x}{\beta(x)^2} \D x=\infty$, 
then $\tau^{\widetilde{S}}=\infty$ almost surely, 
and~\eqref{eq:hitting_equivalence} yields 
$\PP(\tau<\infty) = \PP(\tau^{\widetilde{S}}<\Af_{\infty}) = 0$, 
whatever the value of~$\Af_{\infty}$ and Theorem~\ref{thm:feller}(i) follows.
If $\int_{0}^{1} \frac{x}{\beta(x)^2} \D x<\infty$, 
then $\tau^{\widetilde{S}}<\infty$ almost surely, 
so that, up to a null set, 
$\{\Af_{\infty}=\infty\}\subseteq\{\tau^{\widetilde{S}}<\Af_{\infty}\} = \{\tau<\infty\}$
by~\eqref{eq:hitting_equivalence}, which proves Theorem~\ref{thm:feller}(ii).
The last statement of the theorem follows immediately from~(i)-(ii).
\end{proof}

\subsection{Proof of Theorem~\ref{thm:feeling_bound}}
Since
$\xi_{0}(\cdot) \in \Cc([0,\infty)\to(0,\infty))$, define
\begin{equation}\label{eq:xibar}
0<\underline{\xi}
:= \inf_{t \in [0,1]}\xi_{0}(t)
\leq \sup_{t \in [0,1]}\xi_{0}(t)
=: \overline{\xi}<\infty,
\end{equation}
which do not depend on~$T$, which we assume to belong to $(0,1]$ from now on.
From the continuity assumption on~$\xi_{0}(\cdot)$, 
then~$\Af_{T}$ is finite almost surely for all $T>0$
and furthermore clearly strictly positive since $\sigma_{t}>0$.

\subsubsection{Proof of Theorem~\ref{thm:feeling_bound} (upper bound)}\label{sec:Proof_UpperBound}
We first prove here that
\begin{equation}\label{eq:UpperBound}
\limsup_{T\downarrow 0}\frac{T^{2H}}{\log(T)^2}\log\PP(\tau \leq T)
\leq -\frac{(1-2H)^2}{2\eta^2}.
\end{equation}
By Lemma~\ref{lem:time_change}, \eqref{eq:X_SDE} admits a unique strong solution up to hitting zero.
As in the classical case above, Dambis-Dubins-Schwarz again ensures that there exists a Brownian motion~$\widetilde{W}$ such that
$\int_0^t \sigma_s \D W_s = \widetilde{W}_{\Af_{t}}$, 
with~$\Af_{t}$ in~\eqref{eq:TotalVar}.
The time-changed
process $R_{t} = X_{\Af^{\leftarrow}_{t}}$ (extended to $[0,\infty)$ as in Section~\ref{sec:Bessel_Structure})
solves the Bessel SDE~\eqref{eq:bessel_SDE} with 
$R_0 =
x_{0}$ and hitting time~$\clockhat$,
and 
$\PP(\tau \leq T) = \PP(\clockhat \leq \Af_{T})$
by~\eqref{eq:hitting_equiv}.
For any $m_{T}>0$, write
\begin{align}\label{eq:union_bound}
\PP(\tau \leq T) = \PP\left(\clockhat \leq \Af_{T}\right)
 & = \PP\left(\clockhat \leq \Af_{T}, \Af_{T}\leq m_{T}\right) + \PP\left(\clockhat \leq \Af_{T}, \Af_{T}> m_{T}\right)\nonumber\\
 & \leq \PP\left(\clockhat \leq m_{T}\right) + \PP\left(\Af_{T} > m_{T}\right) =: \PP_{1} + \PP_{2}.
\end{align}
The process~$R$ depends only on
the Brownian motion~$\widetilde{W}$,
so, by Lemma~\ref{lem:bessel_cdf},
using $\Gamma(\nu, x) \leq \mathfrak{C}_{\nu}  x^{\nu - 1}
\E^{-x}$ for~$x$ large enough, $\nu> 0$ and some constant $\mathfrak{C}_{\nu}\geq 1$
\cite[\S6.5.32]{abramowitz1966handbook},
$$
\PP\left(\clockhat \leq m_{T}\right)
= \frac{\Gamma\left(1 - \frac{\delta}{2}, 
\frac{x_{0}^2}{2m_{T}}\right)}{\Gamma\left(1 - \frac{\delta}{2}\right)}
\leq
\frac{\mathfrak{C}_{1 - \frac{\delta}{2}}  \left(\frac{x_{0}^2}{2}\right)^{-\frac{\delta}{2}}}{\Gamma\left(1 - \frac{\delta}{2}\right)} m_{T}^{\frac{\delta}{2}}\exp\left\{-\frac{x_{0}^2}{2m_{T}}\right\}.
$$
Let $m_{T} = \overline{\xi} T \E^{u_{T}}$ for some function~$u_{T}$
such that $\lim_{T\downarrow 0}m_{T} = 0$
and $\lim_{T\downarrow 0}u_{T} = \infty$. 
At leading order,
\begin{equation}\label{eq:T2H_P1}
T^{2H}\log(\PP_{1})
\leq -\frac{x_{0}^2}{2\overline{\xi}}\frac{T^{2H-1}}{\E^{u_{T}}}
+ \Oo\left(T^{2H}\left[1+u_{T}+|\log(T)|\right]\right).
\end{equation}
Now, $\Af_{T}
\leq \int_0^T\xi_0(t)\D t
\E^{\sup_{t\in[0,T]}M_t}
\leq \overline{\xi}T\E^{\sup_{t\in[0,T]}M_t}$,
so that
$$
\PP_2
\leq \PP\left(\sup\limits_{t\in[0,T]}M_{t} > \log\left(\frac{m_T}{\overline{\xi}T}\right)\right)
\leq \exp\left\{
-\frac{\left(\log\left(\frac{m_T}{\overline{\xi}T}\right)-\mu\right)^2}{2\sup_{t\in[0,T]}\VV[M_{t}]}
\right\},
$$ 
by Borell-TIS inequality~\cite[Theorem~2.1.1]{adler2007random}, with $\mu:=\EE\left[\sup_{t\in[0,T]}M_{t}\right] = T^{H}\EE\left[\sup_{t\in[0,1]}M_{t}\right]$ by self-similarity,
as soon as $\log\left(\frac{m_T}{\overline{\xi}T}\right)-\mu>0$, that is for~$T$ small enough.
Since $\sup_{t\in[0,T]}\VV[M_t]=\eta^2T^{2H}$ and~$\mu$ is finite, then, as $T\downarrow 0$,
\begin{equation}\label{eq:T2H_P2}
T^{2H}\log(\PP_{2})
\leq -\frac{u_{T}^2(1+o(1))}{2\eta^2}.
\end{equation}
The sharpest bound from~\eqref{eq:T2H_P2} alone, optimised
over $u_{T}$, is obtained by making $u_{T}$ as large as possible subject to~$\PP_{1}$ remaining
negligible, namely, 
as $T\downarrow 0$,
\begin{equation}\label{eq:negligibility_condition}
\frac{x_{0}^2}{2\overline{\xi}}
\frac{T^{2H-1}}{\E^{u_{T}}} \geq \frac{u_{T}^2}{2\eta^2}(1+o(1)).
\end{equation}
Set $\af := \frac{x_{0}^2}{2\overline{\xi}}$.
Dropping lower-order terms, the condition $T^{2H}\log(\PP_{1}) = T^{2H}\log(\PP_{2})$
reads
\begin{equation}\label{eq:balance_raw}
\frac{\af T^{2H-1}}{\E^{u_{T}}} = \frac{u_{T}^2}{2\eta^2},
\end{equation}
which can be solved explicitly as
$u_{T}^* = 2\Ww\left(\frac{\eta\sqrt{\af}}{\sqrt{2}}T^{H-\half}\right)$,
where $\Ww$ is the Lambert-$\Ww$ function, and admits the expansion
\begin{equation}\label{eq:u_star_leading}
u_{T}^*
= (2H-1)\log(T)
- 2\log\log\left(\frac{1}{T}\right)
+ \Oo(1),
\end{equation}
as $T\downarrow 0$.  
In particular, 
$u_{T}^* \sim (2H-1)\log(T)$.

\begin{proposition}\label{prop:optimised}
Set $m_{T}^* := \E^{u_{T}^*}\overline{\xi} T$.  
Then
\begin{equation}\label{eq:optimised_bound}
\limsup_{T\downarrow 0} 
\frac{T^{2H}}{\log(T)^2}
\log\PP(\tau\leq T)
\leq -\frac{(1-2H)^2}{2\eta^2}.
\end{equation}
\end{proposition}

\begin{proof}
With $m_{T}^* = \E^{u_{T}^*}\overline{\xi} T$, 
then
$\log(\frac{m_{T}^*}{\overline{\xi} T}) = u_{T}^*$, the Borell-TIS bound gives
\begin{equation}\label{eq:P2_at_Mstar}
T^{2H}\log(\PP_{2})
\leq -\frac{(u_{T}^*)^2(1+o(1))}{2\eta^2}.
\end{equation}
From~\eqref{eq:balance_raw} at $u_{T} = u_{T}^*$,
$\frac{\af T^{2H-1}}{\E^{u_{T}^*}}
= \frac{(u_{T}^*)^2}{2\eta^2}$.
Now~\eqref{eq:T2H_P1} implies that  
$T^{2H}\log(\PP_{1})
\leq -\frac{(u_{T}^*)^2}{2\eta^2} + \Oo(T^{2H}\log(T))$,
and both terms are thus of the same order at~$m_{T}^*$.
Since $\log(\PP_{1}+\PP_{2}) \leq \log(2) + \max\{\log(\PP_{1}),\log(\PP_{2})\}$,
$$
\frac{T^{2H}}{\log(T)^2}\log(\PP_{1} + \PP_{2})
\leq
\frac{T^{2H}\log(2)}{\log(T)^2}
 + \frac{T^{2H}}{\log(T)^2}\max\{\log(\PP_{1}),\log(\PP_{2})\}
$$
From~\eqref{eq:u_star_leading}, 
$(u_{T}^*)^2 =(1-2H)^2 \log(T)^2 (1+o(1))$; 
taking $\limsup_{T\downarrow 0}$ finishes the proof.
\end{proof}

Note that the computations above also show that the optimal truncation level satisfies
$$
m_{T}^* = \frac{2\af\eta^2 \overline{\xi} T^{2H}}{(u^*_{T})^2}
= \Oo\left(\frac{T^{2H}}{\log(T)^2}\right).
$$

\subsubsection{Proof of Theorem~\ref{thm:feeling_bound} (lower bound)}\label{sec:Proof_LowerBound}
We now prove the sharp lower bound

\begin{equation}\label{eq:lower_bound_sharp}
\liminf_{T\downarrow 0}\frac{T^{2H}}{\log(T)^2}\log\PP(\tau \leq T)
\geq -\frac{(1-2H)^2}{2\eta^2}.
\end{equation}
As in Section~\ref{sec:Proof_UpperBound}, all statements will be made for $T\in(0,1]$.
The proof of~\eqref{eq:lower_bound_sharp} relies on the following, 
with~$M_{t}$ defined right after~\eqref{eq:alpha_explicit}:

\begin{lemma}\label{lem:increment}
The inequality
$\VV[M_{t}-M_s]\leq c_{H} \eta^2 (t-s)^{2H}$ holds for all $0\leq s\leq t$
with finite constant
$c_{H} = 1 + 2H \int_{0}^{\infty}|u^{H-\half}-(1+u)^{H-\half}|^2\D u$.
\end{lemma}
\begin{proof}
By It\^o isometry,
for $0\leq s\leq t$,
$$
\VV[M_{t}-M_s]
=2H \eta^2 \left[\int_0^s\left((t-u)^{H-\half}-(s-u)^{H-\half}\right)^2\D u
+\int_s^t (t-u)^{2H-1}\D u\right].
$$
The second integral equals 
$\frac{1}{2H}(t-s)^{2H}$. 
For the first, with $v=s-u$ and 
$v=(t-s)w$,
$$
\int_0^s\left|(v+(t-s))^{H-\half}-v^{H-\half}\right|^2\D v
=(t-s)^{2H}\int_0^{\frac{s}{t-s}}\left|(w+1)^{H-\half}-w^{H-\half}\right|^2\D w,
$$
and the integral is finite.
Bounding the integral to the same over $[0,\infty)$,
the lemma follows by checking integrability directly.
\end{proof}

\begin{lemma}\label{lem:chi}
Recall the covariance~$\CC(\cdot,\cdot)$ from~\eqref{eq:c_def}.
The map $\chi:[0,1]\ni\theta \to\frac{\CC(\theta T,T)}{\eta^2 T^{2H}} \in [0,1]$ satisfies
$\half(1-\theta)^{2H} \leq 1-\chi(\theta) \leq \frac{1+c_{H}}{2} (1-\theta)^{2H}$ for all $\theta\in[0,1]$.
\end{lemma}

\begin{proof}
By self-similarity, of~$M$,
the function~$\chi$ does not depend on~$T$.
Since
$
\VV[M_{T} - M_{t}]
 = \VV[M_{T}] +\VV[M_{t}]
 - 2\CC(t,T)
= \eta^{2}T^{2H} + \eta^{2}t^{2H}
- 2\CC(t,T)$,
then, with $t=\theta T$,
$$
\eta^{2}T^{2H} - \CC(\theta T,T)
 = \frac{\VV[M_{T} - M_{\theta T}] + \eta^2 T^{2H}\left(1 - \theta^{2H}\right)}{2}.
$$
and therefore 
$$
1-\chi(\theta)
= \frac{\eta^2 T^{2H} - \CC(\theta T,T)}{\eta^2 T^{2H}}
= \frac{\VV[M_{T} - M_{\theta T}]}{2\eta^2 T^{2H}}
+ \frac{1 - \theta^{2H}}{2}.
$$
The upper bound then follows from Lemma~\ref{lem:increment} and
the trivial inequality $1-\theta^{2H} \leq (1-\theta)^{2H}$,
and the lower bound from Ito's isometry:
$$
\VV[M_{T} - M_{\theta T}]
 = 2H\eta^{2}\int_{0}^{\theta T}\left((T-u)^{H-\half} - (\theta T-u)^{H-\half}\right)^2\D u
+ \eta^{2}T^{2H}(1-\theta)^{2H}.
$$
\end{proof}

Rewrite now the dynamics of $X_{t} := \frac{S_{t}^{1-\beta}}{1-\beta}$ in~\eqref{eq:X_def}-\eqref{eq:X_SDE} as
\begin{equation}\label{eq:Xdyn}
X_{t} = x_0 + \gamma\int_0^t \frac{\sigma_s^2}{X_s}\D s + \rho V_{t} + \rhobar V^\perp_{t},
\end{equation}
with
$\gamma:=\frac{\beta}{2(\beta-1)}<0$, $V_{t}:=\int_0^t\sigma_s\D B_s$
and $V^\perp_{t}:=\int_0^t\sigma_s\D B^\perp_s$.

\begin{lemma}\label{lem:laplace}
Let $\Lambda(x):=\int_{0}^{1}\E^{x\chi(\theta)}\D \theta$ and
$J(x):=\int_{0}^{1} \E^{\frac{x}{2}\chi(\theta)}(1-\theta)^{H-\half}\D \theta$,
with~$\chi$ in Lemma~\ref{lem:chi}. 
Then there exists $\mathfrak{C}_{J}>0$
such that $J(x)\leq \mathfrak{C}_{J}x^{-\half}\sqrt{\Lambda(x)}$ for all $x\geq 1$.
\end{lemma}

\begin{proof}
The bounds from Lemma~\ref{lem:chi} give
$x\chi(\theta) = x-x(1-\chi(\theta))$
so that, with $u=1-\theta$
\begin{align*}
J(x)
= \E^{\frac{x}{2}}\int_{0}^{1} \E^{-\frac{x}{2}(1-\chi(\theta))}(1-\theta)^{H-\half}\D \theta
 & = \E^{\frac{x}{2}}\int_{0}^{1} \E^{-\frac{x}{2}(1-\chi(1-u))}u^{H-\half}\D u\\
 & \leq \E^{\frac{x}{2}}\int_{0}^{\infty} \E^{-\frac{x}{4}u^{2H}}u^{H-\half}\D u\\
& = \frac{\E^{\frac{x}{2}}}{2H}
\int_{0}^{\infty} \E^{-z}
\left(\frac{4}{x}\right)^{\frac{H+\half}{2H}} z^{\frac{H+\half}{2H}-1}\D z\\
 & = \frac{4^{\frac{H+\half}{2H}}}{2H}\Gamma\left(\frac{H+\half}{2H}\right)
 \E^{\frac{x}{2}}x^{-\frac{H+\half}{2H}},
\end{align*}
with the change of variable $z=\frac{x}{4}u^{2H}$.
Similarly, for $x\geq 1$,
\begin{align*}
\Lambda(x)
= \E^{x}\int_{0}^{1} \E^{-x(1-\chi(\theta))}\D \theta
& \geq \E^{x}\int_{0}^{1} \E^{-x \overline{c}u^{2H}}\D u\\
& \geq \frac{\E^{x}(x\overline{c})^{-\frac{1}{2H}}}{2H}\int_{0}^{x\overline{c}} \E^{-z}z^{\frac{1}{2H}-1}\D z\\
& \geq \E^{x}x^{-\frac{1}{2H}}\frac{\overline{c}^{-\frac{1}{2H}}}{2H}\int_{0}^{\overline{c}} \E^{-z}z^{\frac{1}{2H}-1}\D z
= \E^{x}x^{-\frac{1}{2H}}\frac{\overline{c}^{-\frac{1}{2H}}}{2H}\gamma\left(\frac{1}{2H},\overline{c}\right),
\end{align*}
with the change of variable $z = \overline{c}xu^{2H}$ and $\overline{c}:=\frac{1+c_{H}}{2}$,
where~$\gamma(\cdot)$ is the lower incomplete Gamma function.
The bound~$\mathfrak{C}_{J}$ for $\frac{J(x)}{\sqrt{\Lambda(x)}}$ is then immediate, and in fact explicit.
\end{proof}

Let $c_{T}>0$,
\begin{equation}\label{eq:h_Hf}
h_u := \frac{c_{T}\sqrt{2H}}{\eta T^{2H}} (T-u)^{H-\half} \ind_{[0,T]}(u)
\qquad\text{and}\qquad
\Hf_{T} := \half\int_0^T h_u^2\D u = \frac{c_{T}^2}{2\eta^2 T^{2H}},
\end{equation}
so that
$g_{t}:=\eta\sqrt{2H}\int_0^t (t-u)^{H-\half}h_u\D u
=\frac{c_{T}}{\eta^2 T^{2H}} \CC(t,T) = c_{T} \chi(\frac{t}{T})\ge0$,
with~$\chi$ in Lemma~\ref{lem:chi} and $g_{T}=c_{T}$.
Let now $L_{T}:=\exp\{\int_0^T h_u\D B_u-\Hf_{T}\}$ and 
$\D \widetilde{\PP}:=L_{T}\D \PP$. 
Under~$\widetilde{\PP}$, $\widetilde{B} := B -\int_0^{\cdot} h_u\D u$ is a Brownian motion
and $\widetilde{M}_{t}:=\eta\sqrt{2H}\int_0^t (t-u)^{H-\half}\D \widetilde{B}_u$
has the same law under~$\widetilde{\PP}$ as~$M$ under~$\PP$ and $M = \widetilde{M} + g$, $\widetilde{\PP}$-almost surely. 
Now,
$L_{T}^{-1} =\exp\{-H_{T}-\Hf_{T}\}$,
with $H_{T}:=\int_0^T h_u\D \widetilde{B}_u\sim\Nn(0,2\Hf_{T})$
under~$\widetilde{\PP}$. 
Therefore, 
\begin{equation}\label{eq:cm}
\PP(\Af_{T}\geq m_{T})
 = \widetilde{\EE}\left[L_{T}^{-1} \ind_{\Af_{T}\geq m_{T}}\right]
 = \widetilde{\EE}\left[\E^{-H_{T}-\Hf_{T}} \ind_{\Af_{T}\geq m_{T}}\right].
\end{equation}
Under~$\widetilde{\PP}$, 
$\Af_{T}
 = Q_{T}\int_0^T \E^{\widetilde{M}_{t}}\D \nu_{T}(t)$
with
\begin{equation}\label{eq:q_Q}
q_{t} := \xi_{0}(t)\exp\left\{g_{t}-\frac{\eta^2}{2} t^{2H}\right\}
\qquad\text{and}\qquad
Q_T:=\int_{0}^{T} q_{t}\D t,
\end{equation}
then
$\D \nu_{T}(t)
:= \frac{q_{t}}{Q_{T}} 
\D t$
is a probability measure on $[0,T]$. 
Since $\widetilde{\EE}[\E^{\widetilde{M}_t}]
= \exp\left\{\frac{\eta^2 t^{2H}}{2}\right\}$
and
$\int_{0}^{T}\E^{c_{T}\chi(\frac{t}{T})}\D t = T\Lambda(c_{T})$, then
\begin{equation}\label{eq:EtildeA}
\underline{\xi}T\Lambda(c_{T})
\leq \widetilde{\EE}[\Af_{T}] = \int_{0}^{T}\xi_{0}(t)\E^{g_{t}}\D t
\leq \overline{\xi}T\Lambda(c_{T})
\end{equation}

\begin{lemma}\label{lem:pert}
Let $u_{T}:=\log(T^{2H-1})$ and~$c_{T}$ solve $\Lambda(c_{T})=\E^{u_{T}}$. 
Then, for small~$T$,
$\widetilde{\EE}[|V_{T}|] \leq \mathfrak{C}_{V}\sqrt{c_{T}}$ 
with $\mathfrak{C}_{V}:=\sqrt{\overline{\xi}}\left(1+\frac{\sqrt{2H}}{\eta}\mathfrak{C}_{J}\right)$
and~$\mathfrak{C}_{J}$ in Lemma~\ref{lem:laplace}.
\end{lemma}

\begin{proof}
The choice of~$c_{T}$ and~$u_{T}$
imply that~$\widetilde{\PP}$ satisfies $\widetilde{\EE}[\Af_{T}] \leq \overline{\xi}T\E^{u_{T}} = \overline{\xi}T^{2H}$by~\eqref{eq:EtildeA}.
Under~$\widetilde{\PP}$, 
$\D B_{s} = \D \widetilde{B}_{s} + h_{s}\D s$, so $V_{T} = \int_0^T\sigma_{s}\D \widetilde{B}_{s} + \int_0^T\sigma_s h_s\D s$. 
The first term is a $\widetilde{\PP}$-martingale, 
and Jensen's inequality and Ito's isometry give, for~$T$ small enough,
$$
\widetilde{\EE}\left[\left|\int_0^T\sigma_{s}\D \widetilde{B}_{s}\right|\right]
\leq \widetilde{\EE}[\Af_{T}]^{\half}
\leq \sqrt{\overline{\xi}}T^{H}
\leq \sqrt{\overline{\xi} c_{T}}.
$$
Indeed, since $\Lambda(c_{T}) \leq \E^{c_{T}}$ from Lemma~\ref{lem:laplace},
then $c_{T}\geq \log\Lambda(c_{T}) = u_{T} = (1-2H)|\log(T)|$, which is greater than~$1$ for~$T$ small enough, while~$T^{2H}\leq 1$ for such~$T$.
For the second term, $h\ge0$ and
$\widetilde{\EE}[\sigma_s]
 = \sqrt{\xi_{0}(s)}\E^{\frac{g_s}{2}-\frac{1}{8}\eta^2 s^{2H}}
 \leq \sqrt{\overline{\xi}} \E^{\frac{g_s}{2}}$, 
so with~$h_s$ in~\eqref{eq:h_Hf},
\begin{align*}
\widetilde{\EE}
\left[\int_0^T\sigma_s h_s\D s\right]
&  \leq \sqrt{\overline{\xi}} \frac{c_{T}\sqrt{2H}}{\eta T^{2H}}
\int_0^T\E^{\frac{g_s}{2}} (T-s)^{H-\half} \D s\\
& = \sqrt{\overline{\xi}} \frac{c_{T}\sqrt{2H}}{\eta}
\frac{J(c_{T})}{T^{H-\half}}
 \leq \mathfrak{C}_{J}\sqrt{\overline{\xi}} \frac{\sqrt{2Hc_{T}}}{\eta}
\frac{\sqrt{\Lambda(c_{T})}}{T^{H-\half}}
= \mathfrak{C}_{J}\sqrt{\overline{\xi}} \frac{\sqrt{2Hc_{T}}}{\eta},
\end{align*}
using $g_s=c_{T} \chi(\frac{s}{T})$,
Lemma~\ref{lem:laplace}
and $\Lambda(c_{T}) = T^{2H-1}$,
and the lemma follows.
\end{proof}

\begin{proof}[Proof of~\eqref{eq:lower_bound_sharp}]
Keep~$u_{T}$ and~$c_{T}$ as in Lemma~\ref{lem:pert} and use~$\widetilde{\PP}$, the control~$h$,
$\Hf_{T}=\frac{c_{T}^2}{2\eta^2 T^{2H}}$,
$H_{T}=\int_0^T h_{u}\D \widetilde{B}_{u}$
and define the random variable
$Y:=\int_0^T \widetilde{M}_{t}\D \nu_{T}(t)$,
which is centred Gaussian under~$\widetilde{\PP}$
with variance
$\mathfrak{v}_{v}^2 := \iint \CC(s,t)\D \nu_{T}(s)\D \nu_{T}(t)\leq \eta^2 T^{2H}$,
therefore $\widetilde{\PP}(Y < -\mathfrak{v}_{v})
 \leq \Phi(-1)$.
Since~$H_{T}\sim \Nn(0,2\Hf_{T})$ under~$\widetilde{\PP}$, 
then $\widetilde{\PP}(H_{T} > \sqrt{2\Hf_{T}}) = \Phi(-1)$.
By Lemma~\ref{lem:pert} and Markov's inequality, $\widetilde{\PP}(|V_{T}|>4\mathfrak{C}_{V}\sqrt{c_{T}})\leq \frac{1}{4}$.
With the set 
$\Ss_{T} := \{Y\geq -\mathfrak{v}_{v}\}\cap\{H_{T}\leq \sqrt{2\Hf_{T}}\}\cap\{|V_{T}|\leq 4\mathfrak{C}_{V}\sqrt{c_{T}}\}$,
we therefore have
$$
\widetilde{\PP}\left(\Ss_{T}^{c}\right)
\leq \widetilde{\PP}\left(Y< -\mathfrak{v}_{v}\right)
+ \widetilde{\PP}\left(H_{T}> \sqrt{2\Hf_{T}}\right)
+ \widetilde{\PP}\left(|V_{T}|> 4\mathfrak{C}_{V}\sqrt{c_{T}}\right)
\leq 2\Phi(-1) + \frac{1}{4},
$$
which is uniform in~$T$.
Since
$\Ss_{T}\in\Ff^B$, then, under~$\PP$ 
$$
\PP(\Ss_{T})=\widetilde{\EE}\left[L_{T}^{-1}\ind_{\Ss_{T}}\right]
\geq \E^{-\sqrt{2\Hf_{T}}-\Hf_{T}} \widetilde{\PP}(\Ss_{T})
\geq \frac{\exp\left\{-\sqrt{2\Hf_{T}}-\Hf_{T}\right\}}{4}.
$$
From~\eqref{eq:q_Q} and Jensen's inequality,
then, under~$\widetilde{\PP}$,
$\Af_{T}  = Q_{T}\int_0^T \E^{\widetilde{M}_{t}}\D \nu_{T}(t)
\geq Q_{T} \E^{Y}$ and
$Q_{T} = \int_{0}^{T} q_{t}\D t
\geq \underline{\xi}\E^{\frac{-\eta^2 T^{2H}}{2}}T\Lambda(c_{T})
= \underline{\xi}\E^{\frac{-\eta^2 T^{2H}}{2}}T^{2H}$
and therefore, on~$\Ss_{T}$,
since $Y\geq -\mathfrak{v}_{v} \geq -\eta T^{H}$ there,
$\Af_{T} \geq \underline{\xi} T^{2H} 
\exp\left\{-\frac{\eta^2}{2}T^{2H}-\eta T^{H}\right\}
\geq \frac{\underline{\xi}}{2} T^{2H}$,
for~$T$ small enough, which we assume now.
Finally, on $\Ss_{T}$, since 
$|V_{T}|\leq 4\mathfrak{C}_{V}\sqrt{c_{T}}$,
then
\begin{equation}\label{eq:def_zT}
\frac{x_0+\rho V_{T}}{\rhobar \sqrt{\Af_{T}}}
\leq z_{T} := \sqrt{2}\frac{x_0 + 4|\rho| \mathfrak{C}_{V}\sqrt{c_{T}}}{\rhobar \sqrt{\underline{\xi}}T^H}
\end{equation}
and by Lemma~\ref{lem:rSABR_access},
$\PP(\tau\leq T) = 
\EE\left[\PP\left(\tau\leq T| \Ff^B\right)\right]
\geq 
\EE\left[\ind_{\Ss_{T}}\PP\left(\tau\leq T | \Ff^B\right)\right]
 \geq 
\Phi(-z_{T})\PP(\Ss_{T})$.
Because $\Lambda(c_{T}) = \E^{u_{T}}$ and
$\lim\limits_{c\uparrow\infty}\frac{c}{\log(\Lambda(c))} = 1$
using Lemma~\ref{lem:chi}, then $c_{T}=u_{T}(1+o(1))$ and
$z_{T}^2 = \Oo\left(c_{T}T^{-2H}\right) = \Oo\left(|\log(T)|T^{-2H}\right)$ and therefore
$$
\log\Phi(-z_{T}) \geq -\frac{z_{T}^{2}}{2}-\log\left(2z_{T}\sqrt{2\pi}\right)
 = \Oo\left(\frac{|\log(T)|}{T^{2H}}\right)
 = o\left(\frac{\log(T)^2}{T^{2H}}\right),
$$
using the Gaussian tail
$\Phi(-z)\geq \frac{1}{2z\sqrt{2\pi}}\exp\{-\frac{z^2}{2}\}$ for $z\geq 1$.
Since $\sqrt{2\Hf_{T}} = o(\Hf_{T})$, then
$$
\log\PP(\Ss_{T}) \geq -\Hf_{T}(1+o(1))
= - \frac{c_{T}^2}{2\eta^2T^{2H}}(1+o(1))
= - \frac{(1-2H)^2\log(T)^2}{2\eta^2T^{2H}}(1+o(1)).
$$
Dividing by $\frac{\log(T)^2}{T^{2H}}$ and taking $\liminf$ proves the claim.
\end{proof}


\subsection{Proof of Theorem~\ref{thm:feeling_bound_general}}

Assumption~\ref{ass:beta} is in force,
$\underline{\xi}$ and~$\overline{\xi}$ 
given in~\eqref{eq:xibar} and $T\in(0,1)$.
From Lemma~\ref{lem:time_change}, pathwise uniqueness holds for~\eqref{eq:S_SDE} up to~$\tau$, 
and~$S$ does not explode since $S_{t\wedge\tau}=\widetilde{S}_{\Af_{t}}$.
With~$\varphi$ in Assumption~\ref{ass:beta}, define, consistently with~\eqref{eq:X_def},
$$
X_{t} := \varphi(S_{t})
\qquad\text{and}\qquad
x_{0} := \varphi(S_{0})>0,
$$
and~$\varphi$ is a smooth increasing bijection from $(0,\infty)$ to $(0,\varphi(\infty))$,
with $\lim_{s\downarrow 0}\varphi(s)=0$, $\varphi'(s) = \frac{1}{\beta(s)}$, $\varphi''(s) = -\frac{\beta'(s)}{\beta(s)^2}$,
so that $\tau = \inf\{t\geq 0: X_{t}=0\}$ and It\^o's formula yields
\begin{equation}\label{eq:X_SDE_general}
\D X_{t}
=  - \frac{\beta'(S_{t})}{2}\sigma_{t}^2\D t
+ \sigma_{t}\D W_{t}
=  - \frac{\beta'(S_{t})}{2}\sigma_{t}^2\D t 
+ \rho \D V_{t} + \rhobar\D V^{\perp}_{t},
\end{equation}
for $t<\tau$,
with~$V_{T}$ and~$V^{\perp}_{T}$ as defined previously (after~\eqref{eq:Xdyn}).
In contrast with the power-law case from Theorem~\ref{thm:feeling_bound},
the time-changed process is no longer Bessel and the explicit formula in Lemma~\ref{lem:bessel_cdf} is not available,
but is replaced by the estimate in Lemma~\ref{lem:hitting_general}.
We start with the analogue of Lemma~\ref{lem:rSABR_access}:

\begin{lemma}\label{lem:condhit_general}
For every $T>0$, 
Lemma~\ref{lem:rSABR_access} holds with $x_{0} = \varphi(S_{0})$.
\end{lemma}

\begin{proof}
Let $T>0$.
On $\{\tau>T\}$, 
$S$ is continuous and strictly positive on $[0,T]$,
so that $\int_{0}^{T}\beta'(S_{s})\sigma_{s}^2\D s$ is finite and non-negative almost surely
and thus
$X_{T} \in (0, x_{0} + \rho V_{T} + \rhobar V^{\perp}_{T}]$ using~\eqref{eq:X_SDE_general}.
Thus
$\{\rhobar V^\perp_{T}\le-(x_0+\rho V_{T})\}\subseteq\{\tau\leq T\}$
and the lemma follows as in the proof of Lemma~\ref{lem:rSABR_access},
since $V^\perp_{T}|\Ff^B\sim\Nn(0,\Af_{T})$ and~$V_{T}$ and~$\Af_{T}$ are $\Ff^{B}$-measurable.
\end{proof}

We now provide a bound on the hitting time 
$\widetilde{\tau} := \inf\{t\geq 0:\widetilde{S}_{t}=0\}$
of the origin by the time-homogeneous diffusion~$\widetilde{S}$ of Lemma~\ref{lem:time_change},
mimicking Lemma~\ref{lem:bessel_cdf}.

\begin{lemma}\label{lem:hitting_general}
Let $\df_{\eps}(m):=(1-\eps)x_{0}-\cf_{\eps}m$ with
$\cf_{\eps} := \half\max\left\{\beta'(s): \eps x_{0}\leq \varphi(s)\leq x_{0}\right\}\in[0,\infty)
$.
Then, for all $\eps\in(0,1)$ and all $m>0$ such that $\df_{\eps}(m)>0$,
\begin{equation}\label{eq:hitting_two_sided}
2\Phi\left(-\frac{x_{0}}{\sqrt{m}}\right)
\leq \PP\left(\widetilde{\tau}\leq m\right)
\leq 4\Phi\left(-\frac{\df_{\eps}(m)}{\sqrt{m}}\right)
\leq 2\exp\left\{-\frac{\df_{\eps}(m)^2}{2 m}\right\}.
\end{equation}
In particular,
$\lim_{m\downarrow 0} m\log\PP\left(\widetilde{\tau}\leq m\right) = -\frac{x_{0}^2}{2}$.
\end{lemma}

\begin{proof}
Applying It\^o's formula to $\widetilde{X}:=\varphi(\widetilde{S})$ gives 
\begin{equation}\label{eq:Xtilde_general}
\widetilde{X}_{t} = x_{0} + \widetilde{W}_{t} - \half\int_{0}^{t}\beta'(\widetilde{S}_{s})\D s,
\qquad\text{for }t<\widetilde{\tau},
\end{equation}
and 
$\lim_{t\uparrow\widetilde{\tau}}\widetilde{X}_{t}=0$ almost surely on 
$\{\widetilde{\tau}<\infty\}$ by assumption.
The lower bound in~\eqref{eq:hitting_two_sided} is immediate:
since~$\beta$ is non-decreasing, \eqref{eq:Xtilde_general} gives
$\widetilde{X}_{t}\leq x_{0}+\widetilde{W}_{t}$ for $t<\widetilde{\tau}$.
On $\{\widetilde{\tau}>m\}$, then
$\widetilde{X}_{t}>0$ for all $t\in[0,m]$,
so that $\{\inf_{t\leq m}\widetilde{W}_{t}\leq -x_{0}\}\subseteq\{\widetilde{\tau}\leq m\}$
and the reflection principle for Brownian motion~\cite[\S2]{karatzas1998brownian} yields
$$
\PP\left(\widetilde{\tau}\leq m\right)
\geq \PP\left(\inf_{t\leq m}\widetilde{W}_{t}\leq -x_{0}\right)
= 2\Phi\left(-\frac{x_{0}}{\sqrt{m}}\right).
$$
Fix now $\eps\in(0,1)$, $m>0$ such that $\df_{\eps}(m)>0$ 
and consider the event 
$\{\widetilde{\tau}\leq m\}$.
Since~$\widetilde{X}$ is continuous, then
$\overline{\vartheta}:=\inf\{t\geq 0:\widetilde{X}_{t}=\eps x_{0}\} < \widetilde{\tau}$
and $\underline{\vartheta}:=\sup\{t\leq\overline{\vartheta}:\widetilde{X}_{t}=x_{0}\}\in[0,\overline{\vartheta})$ are well defined.
For $t\in[\underline{\vartheta},\overline{\vartheta}]$, $\widetilde{X}_{t}\in[\eps x_{0},x_{0}]$,
so that $\half\beta'(\widetilde{S}_{t})\in [0,\cf_{\eps}]$,
and~\eqref{eq:Xtilde_general} yields
$$
(\eps-1)x_{0}
= \widetilde{X}_{\overline{\vartheta}}-\widetilde{X}_{\underline{\vartheta}}
= \widetilde{W}_{\overline{\vartheta}}-\widetilde{W}_{\underline{\vartheta}} - \half\int_{\underline{\vartheta}}^{\overline{\vartheta}}\beta'(\widetilde{S}_{s})\D s
\geq \widetilde{W}_{\overline{\vartheta}}-\widetilde{W}_{\underline{\vartheta}} - \cf_{\eps}m.
$$
Hence $\widetilde{W}_{\underline{\vartheta}}-\widetilde{W}_{\overline{\vartheta}}\geq \df_{\eps}(m)$ 
so that
$\{\widetilde{\tau}\leq m\}\subseteq\{D_{m}\geq \df_{\eps}(m)\}$, where
$$
D_{m} := \sup_{0\leq s\leq t\leq m}\left\{\widetilde{W}_{s} - \widetilde{W}_{t}\right\}
 = \sup_{t\leq m}\left\{\sup_{s\leq t}\{\widetilde{W}_{s}\} - \widetilde{W}_{t}\right\}.
$$
By L\'evy's theorem~\cite[\S\,VI, Theorem~2.3]{revuz2013continuous},
$(\sup_{s\leq t}\{\widetilde{W}_{s}\} - \widetilde{W}_{t})_{t\geq 0}$ has the same law as $(|\widetilde{W}_{t}|)_{t\geq 0}$,
so that~$D_{m}$ has the same law as $\sup_{t\leq m}|\widetilde{W}_{t}|$,
and the reflection principle gives
$$
\PP\left(\widetilde{\tau}\leq m\right)
\leq \PP\left(\sup_{t\leq m}\{|\widetilde{W}_{t}|\}\geq \df_{\eps}(m)\right)
\leq 2\PP\left(\sup_{t\leq m}\{\widetilde{W}_{t}\}\geq \df_{\eps}(m)\right)
= 4\Phi\left(-\frac{\df_{\eps}(m)}{\sqrt{m}}\right).
$$
The last inequality in~\eqref{eq:hitting_two_sided} follows from $\Phi(-z)\leq\half\E^{-\frac{z^2}{2}}$ for $z\geq 0$
and the limit statement is straightforward.
\end{proof}

Note that the constant~$\frac{x_{0}^2}{2}$ in the limit in Lemma~\ref{lem:hitting_general} is the same as in Lemma~\ref{lem:bessel_cdf} since $\PP(\clock\leq m) = \exp\{-\frac{x_{0}^2}{2m}(1+o(1))\}$ as $m\downarrow 0$.
We now move on to the actual proof of the theorem.

\begin{proof}[Proof of Theorem~\ref{thm:feeling_bound_general}]
The lower bound~\eqref{eq:lower_bound_sharp} follows the exact same step as in the power law case from Section~\ref{sec:Proof_LowerBound},
with Lemma~\ref{lem:rSABR_access} replaced by Lemma~\ref{lem:condhit_general}:
Lemmas~\ref{lem:increment}, \ref{lem:chi}, \ref{lem:laplace}, \ref{lem:pert},
the change of measure~$\widetilde{\PP}$ and the event~$\Ss_{T}$ only involve the volatility process~$\sigma$ and the Brownian motion~$B$,
while the dynamics of~$S$ only enter through the conditional bound of Lemma~\ref{lem:rSABR_access}
and through the constant~$x_{0}$ in the definition of~$z_{T}$ in~\eqref{eq:def_zT}.
For the upper bound, by Lemma~\ref{lem:bijection}, 
$\{\tau\leq T\} = \{\tau^{\widetilde{S}}\leq\Af_{T}\}$, so that, for any $m_{T}>0$,
$\PP(\tau\leq T)
\leq \PP(\tau^{\widetilde{S}}\leq m_{T}) + \PP(\Af_{T}>m_{T})
=: \PP_{1}+\PP_{2}$.
Take $m_{T}:=\frac{T^{2H}}{|\log(T)|^{3}}$, which tends to zero as $T\downarrow 0$.
The limit in Lemma~\ref{lem:hitting_general} then yields
$$
\frac{T^{2H}}{\log(T)^2}\log(\PP_{1})
= -\frac{T^{2H}}{\log(T)^2}\frac{x_{0}^2}{2m_{T}}(1+o(1))
= -\frac{x_{0}^2}{2}|\log(T)|(1+o(1)).
$$
As in Section~\ref{sec:Proof_UpperBound},
$\Af_{T}\leq\overline{\xi}T\E^{\sup_{t\in[0,T]}M_{t}}$;
by Borell-TIS inequality for~$T$ small enough,
$$
\PP_{2}
\leq \exp\left\{-\frac{1}{2\eta^2T^{2H}}\left[\log\left(\frac{m_{T}}{\overline{\xi}T}\right)-\mu\right]^2\right\},
$$
with $\mu := T^{H}\EE[\sup_{t\in[0,1]}M_{t}]$, giving
$\frac{T^{2H}}{\log(T)^2}\log(\PP_{2})
\leq -\frac{(1-2H)^2}{2\eta^2}(1+o(1))$.
Again $\log(\PP_{1}+\PP_{2}) \leq \log(2) + \max\{\log(\PP_{1}),\log(\PP_{2})\}$ 
and the upper bound and the theorem follow.
\end{proof}

\begin{remark}
The choice $m_{T}=\frac{T^{2H}}{|\log(T)|^{3}}$, combined with Lemma~\ref{lem:bessel_cdf},
provides a direct proof of~\eqref{eq:UpperBound} in the power-law case,
bypassing the optimisation over~$u_{T}$ of Section~\ref{sec:Proof_UpperBound}.
Conversely, this optimisation can be reproduced here,
using Lemma~\ref{lem:hitting_general} instead of Lemma~\ref{lem:bessel_cdf}:
for any $\eps\in(0,1)$, with $m_{T}=\overline{\xi}T\E^{u_{T}}$ tending to zero, \eqref{eq:hitting_two_sided} gives
$T^{2H}\log(\PP_{1})\leq -(1-\eps)^2\af\frac{T^{2H-1}}{\E^{u_{T}}}+\Oo(T^{2H})$,
with the same~$\af$ as in the power-law case.
\end{remark}

\bibliographystyle{siam}
\bibliography{references}
\end{document}